\documentclass[twocolumn,prl,prb,nobibnotes,altaffilletter,amsmath,amssymb,amsfonts]{revtex4}
\date{\today}

\usepackage[latin1]{inputenc}
\usepackage{graphicx}
\usepackage{amsmath}
\usepackage{latexsym}
\usepackage{epsfig}

\begin{document}

\title{Supercooled Liquid Dynamics Studied via Shear-Mechanical Spectroscopy}
\author{Claudio Maggi\footnote{Electronic address: cmaggi@ruc.dk}}
\author{Bo Jakobsen}
\author{Tage Christensen}
\author{N. B. Olsen}
\author{Jeppe C. Dyre}
\affiliation{DNRF Centre 'Glass and Time', IMFUFA, Department of Sciences, Roskilde University, Postbox 260, DK-4000 Roskilde, Denmark}

\begin{abstract}

We report dynamical shear-modulus measurements for five glass-forming liquids (pentaphenyl trimethyl trisiloxane, diethyl phthalate, dibutyl phthalate, 1,2-propanediol, and m-touluidine). The shear-mechanical spectra are obtained by the piezoelectric shear-modulus gauge (PSG) method. This technique allows one to measure the shear modulus ($10^{5} -10^{10}$ Pa) of the liquid within a frequency 
range from 1 mHz to 10 kHz. We analyze the frequency-dependent response functions to investigate whether time-temperature superposition (TTS) is obeyed. We also study the shear-modulus loss-peak position and its high-frequency part. It has been suggested that when TTS applies, the 
high-frequency side of the imaginary part of the dielectric response  decreases like a power law of the frequency with an exponent $-1/2$. This conjecture is analyzed on the basis of the shear mechanical data. We find that TTS is obeyed for pentaphenyl trimethyl trisiloxane and in 1,2-propanediol
while in the remaining liquids evidence of a mechanical $\beta$ process is found. Although the the high-frequency power law behavior $\omega^{-\alpha}$ of the shear-loss may approach a limiting value of $\alpha=0.5$ when lowering the temperature, we find that the exponent lies systematically above this value (around $0.4$). For the two liquids without beta relaxation (pentaphenyl trimethyl trisiloxane and 1,2-propanediol) we also test the shoving model prediction, according to which the the relaxation-time activation energy is proportional to the instantaneous shear modulus. We find that the data are well described by this model.

\end{abstract}

\maketitle

\section{I. Introduction}

The nature of the relaxation processes taking place in supercooled liquids approaching the glass transition has been a major subject of study for a number of years. Understanding how the different response functions are connected in such systems is still a fundamental goal to reach. It is not clear whether the various observables display some universal features approaching the glass transition. Although dielectric spectroscopy is the most common experimental tool, a more detailed characterization of their behavior can be obtained by measuring different quantities, like the \emph{shear modulus}, that are important from a practical as well as theoretical point of view. Although this dynamic variable can be measured above $\sim 10^2$ Hz with conventional techniques, high-frequency data are scarce in the literature.

Motivated by these reasons we employed the piezoelectric-shear-modulus-gauge (PSG)~\cite{RevSci} method to measure the shear modulus of five glass-forming liquids. This technique allows us to measure the dynamic shear modulus of the supercooled liquids just above the glass transition where it takes values between 0.1 MPa and 10 GPa. By means of the PSG technique we can easily observe the $\alpha$ relaxation process in the shear response. The frequency range of the technique is wide ($10^{-3} - 10^{4}$ Hz), and we also observe a \emph{mechanical} Johari-Goldstain $\beta$ relaxation~\cite{beta}. As shown before~\cite{Kr&Bo} indeed, this technique is sensitive to the secondary process and we find evidence of the  presence of a shear $\beta$ relaxation~\cite{Kr&Bo,Bo,DiMarzio} in some of the mechanical spectra reported here.

In Section II we describe the experiment performed and the liquids studied reporting the frequency resolved mechanical spectra. In Sec. III we report the main findings of our study. In Section III we present the analyzed data showing the temperature dependence of the shear-mechanical $\alpha$ relaxation frequency and a test of some conjectures and models about the dynamics. Finally we draw some general conclusions in Sec. IV.

\section{II. Experiment} 
\label{sec:Exp}

\begin{figure}
\begin{center}
\includegraphics[width=8cm]{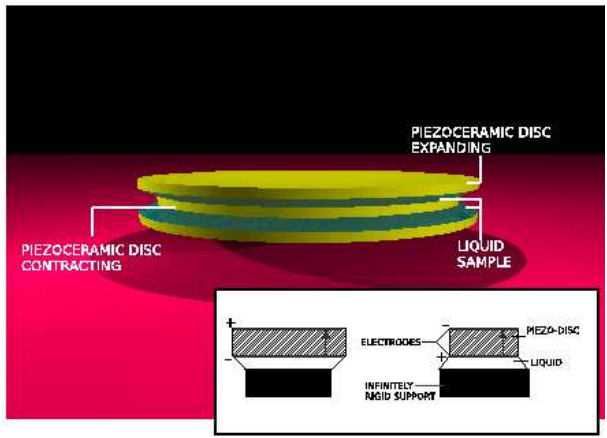}
\end{center}
\caption{(Main panel) Pictorial representation of the shear-transducer (PSG) during its expansion-contraction. The liquid sandwiched between the discs clamps their movement inducing a measurable change in the piezoelectric capacitance of the disc-system. The shear modulus can be deduced from this electrical capacitance. (Inset) The one-disc equivalent system of the PSG. The applied electric field causes an expansion or a contraction depending on the polarization of the disc (represented by the arrow). See Ref.~\cite{RevSci} for more details.}
\label{fig:PSG}
\end{figure}

The piezoelectric shear-modulus gauge (PSG) method is based on the piezoelectric properties of the material that composes the measuring device. The piezoelectric transducer is formed by three discs made of a special ceramic compound~\cite{PZ} that has a pronounced piezoelectric effect (see Fig. \ref{fig:PSG}). The working principle of the PSG is illustrated in the inset of Fig. \ref{fig:PSG} where the one disc equivalent of the three discs system is shown. The ceramic disc is covered with a silver layer on both faces constituting the electrodes. When a voltage is applied, each disc expands or contracts depending on its intrinsic polarization and on the direction of the acting electric field. The electric capacitance of the disc depends on its strain state so that, if a material is partially clamping its motion, the measured capacitance will be lower than that of the free moving disc. Measuring accurately this electrical capacitance~\cite{Br2} we can obtain the stiffness of the medium adhering to the disc. In other words we can convert the electric impedance into the shear modulus knowing the exact relationship between the two~\cite{RevSci}. The three-discs geometry, used in the experiment (main panel of Fig. \ref{fig:PSG}) is employed to reduce unwanted effect like the bending motion that would be present in a one-disc device. The interested reader can find details about the technique in Ref.~\cite{RevSci}.

The measurements are performed cooling the liquids via a home-built closed-cycle cryostat~\cite{Brian}. This has an absolute uncertainty on the temperature that is less than 0.2 K and a temperature stability better than 2 mK.

The liquids studied are the following: pentaphenyl trimethyl trisiloxane (DC705),  dibutyl phthalate (DBP), diethyl phthalate (DEP), 1,2-propanediol, and m-touluidine. The DC705 is a diffusion-pump oil from Dow Corning, all the other liquids were acquired from Sigma-Aldrich. No filtration or purification of the samples was performed before the measurement.

\begin{figure*}
\centering
\begin{tabular}{cc}
\epsfig{file=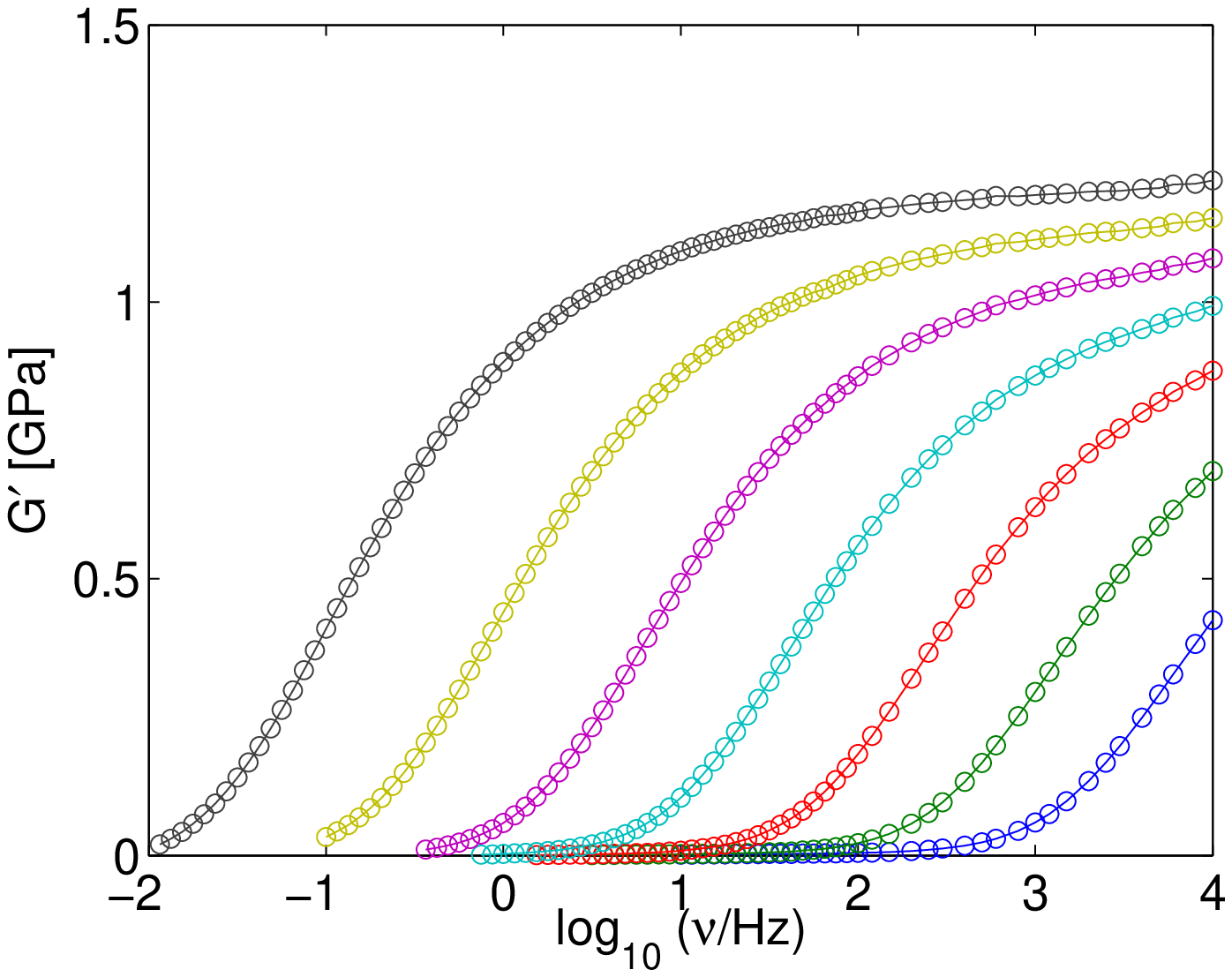,width=0.35\linewidth,clip=} &
\epsfig{file=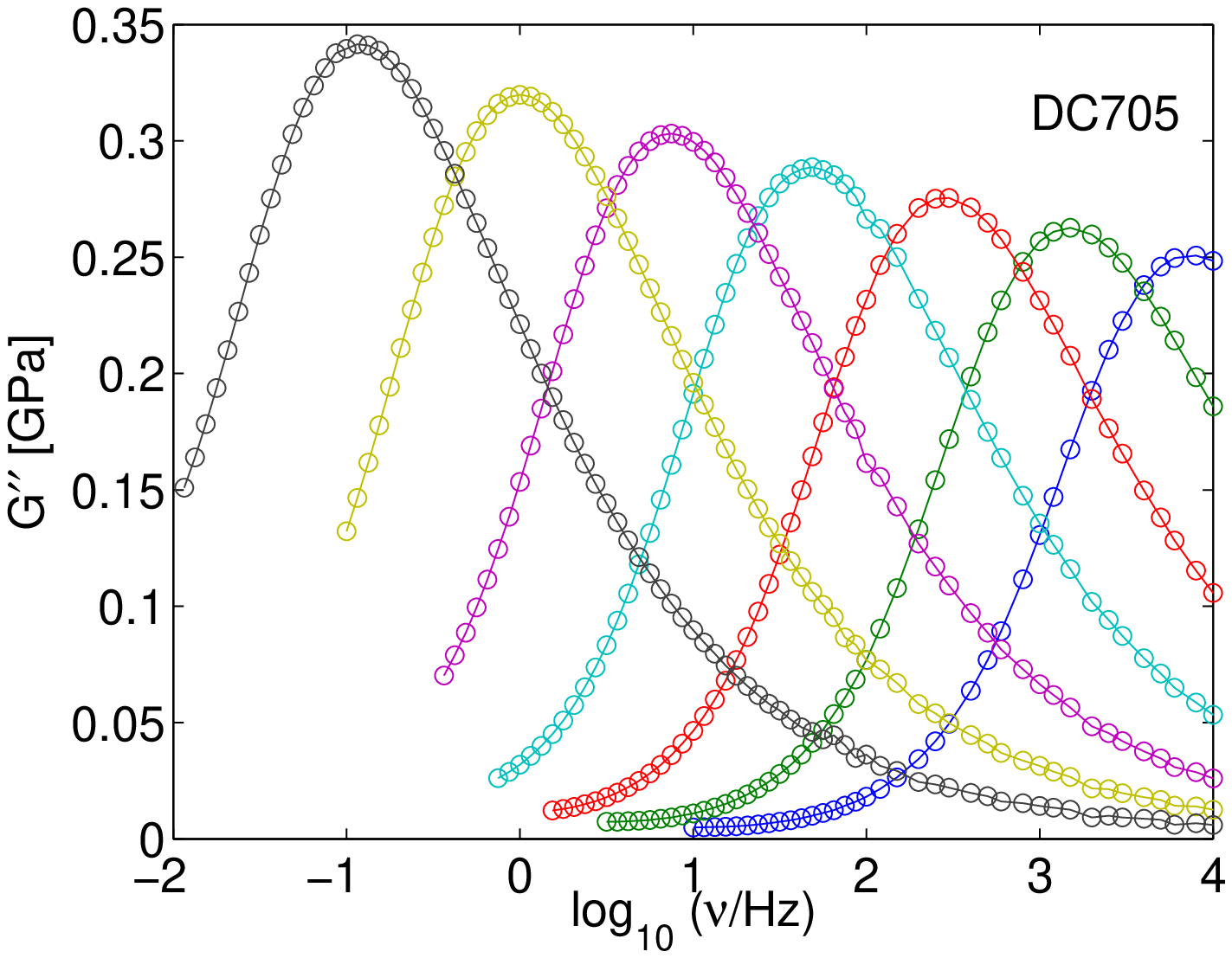,width=0.35\linewidth,clip=} \\
\epsfig{file=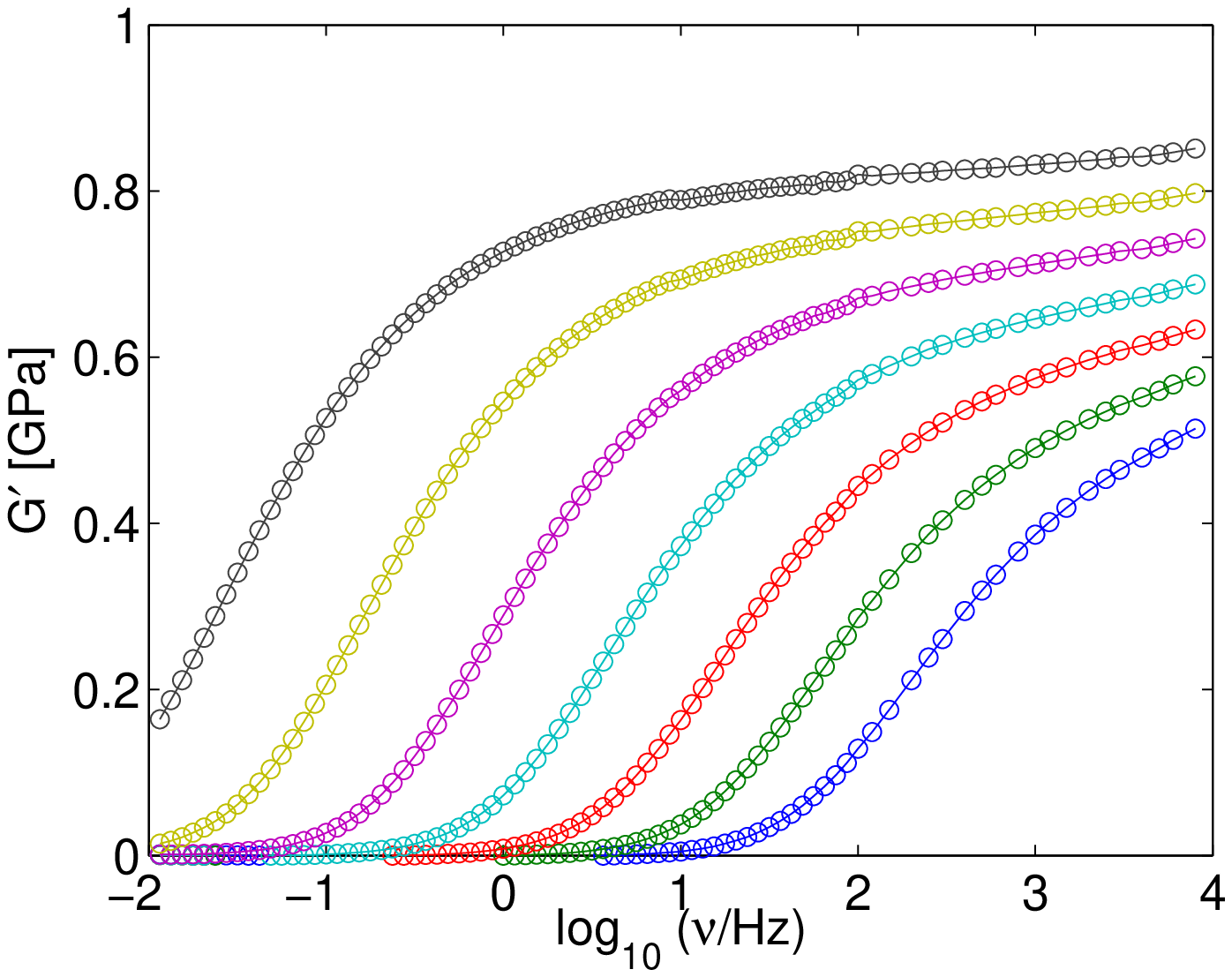,width=0.35\linewidth,clip=} &
\epsfig{file=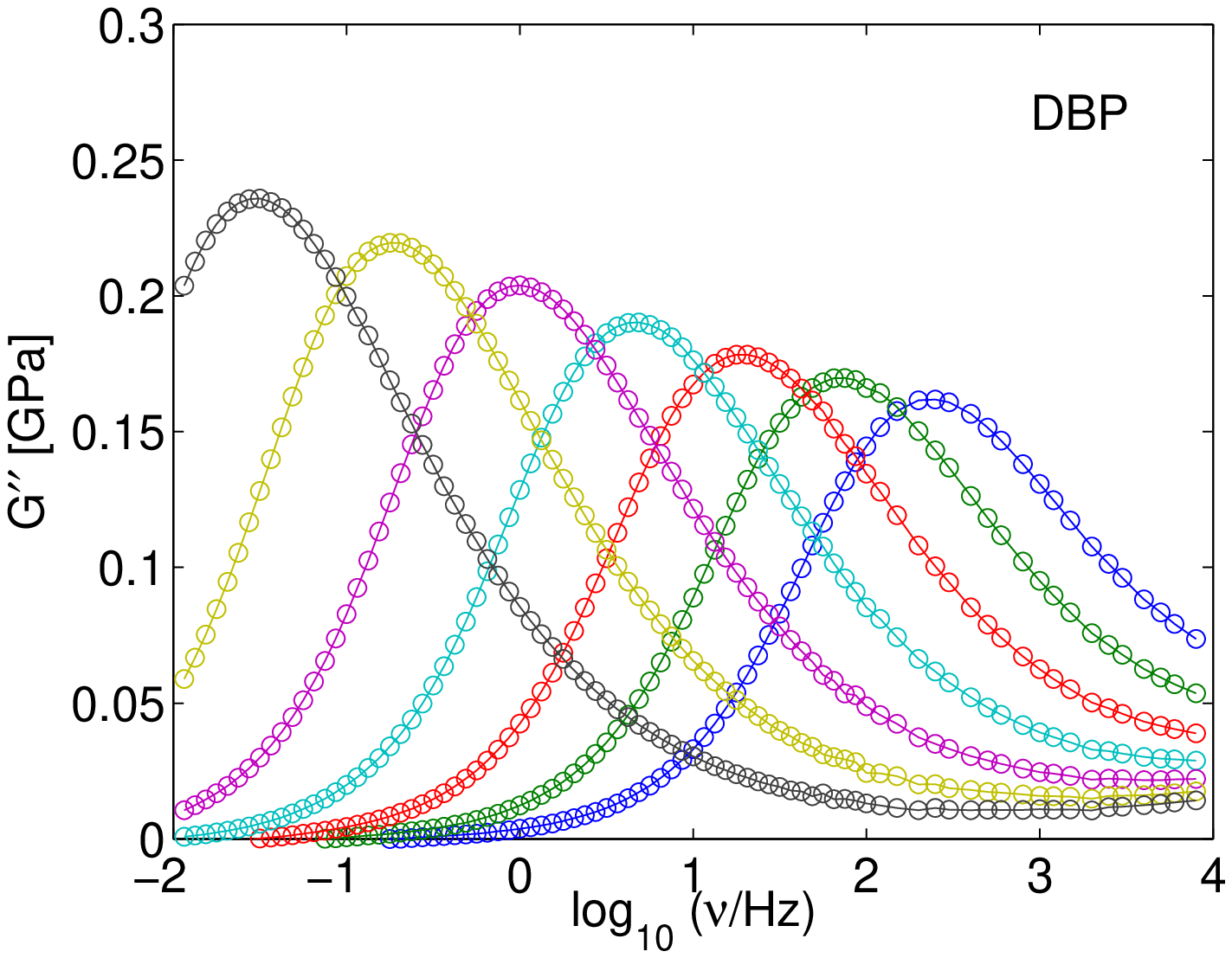,width=0.35\linewidth,clip=} \\
\epsfig{file=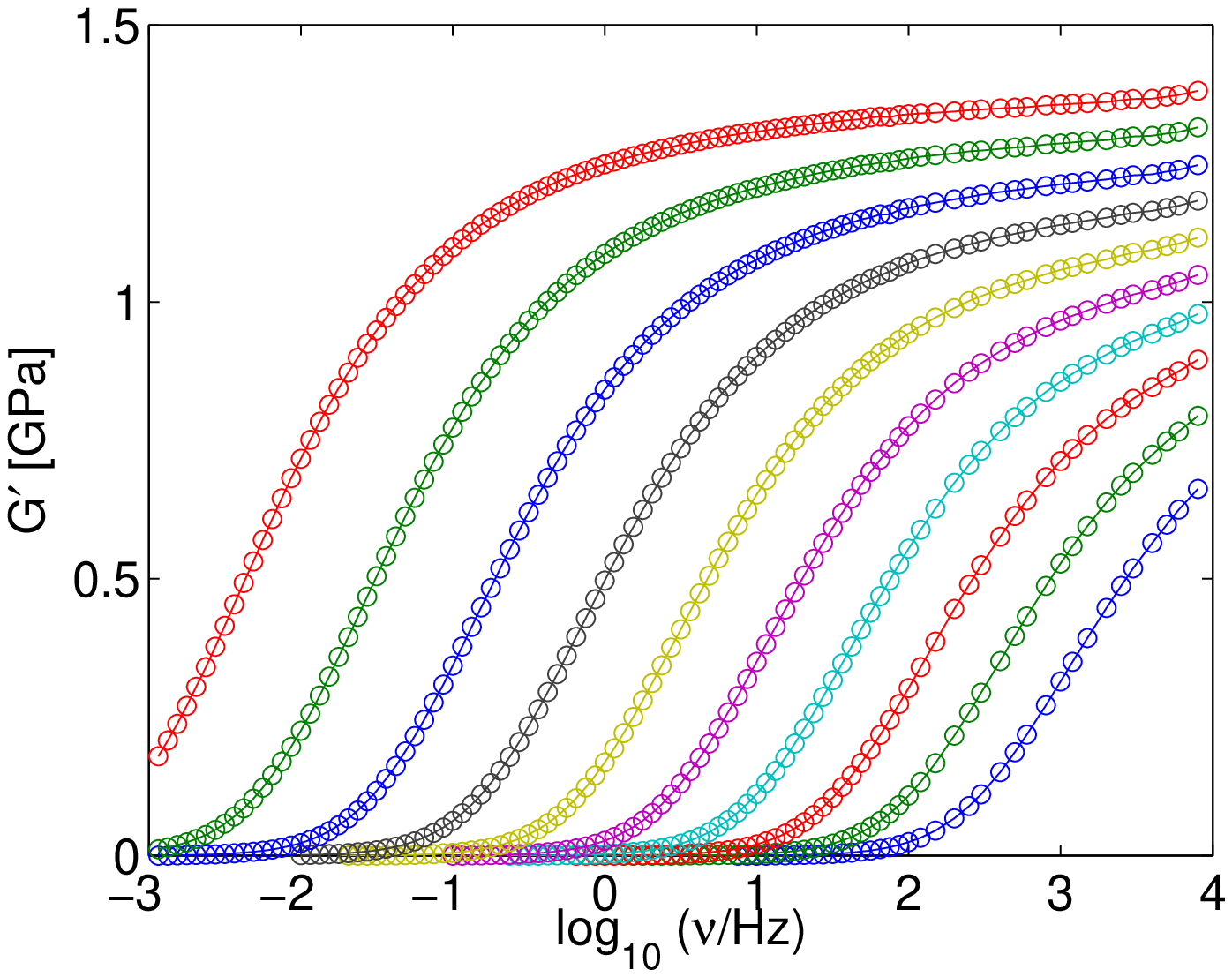,width=0.35\linewidth,clip=} &
\epsfig{file=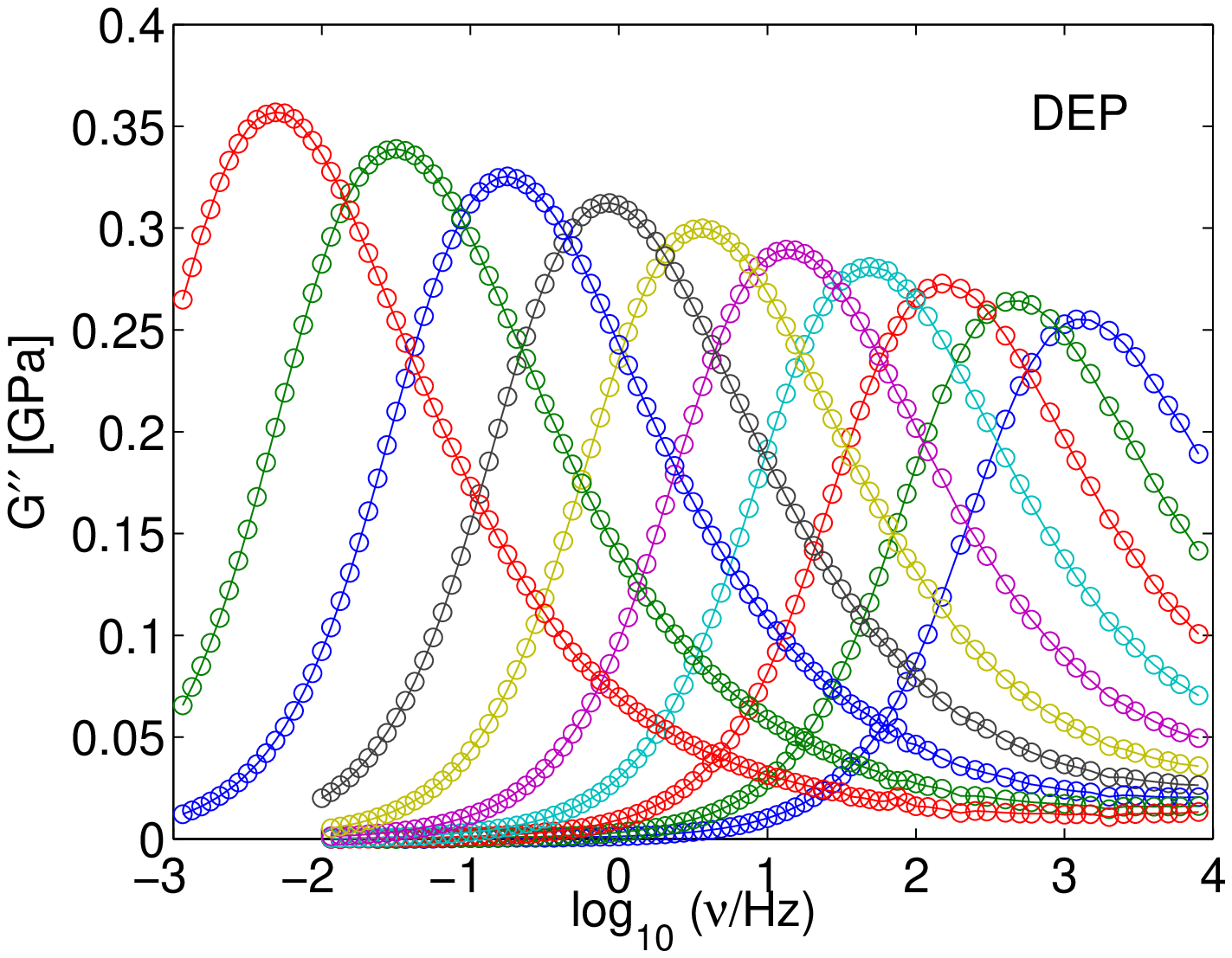,width=0.35\linewidth,clip=} \\
\epsfig{file=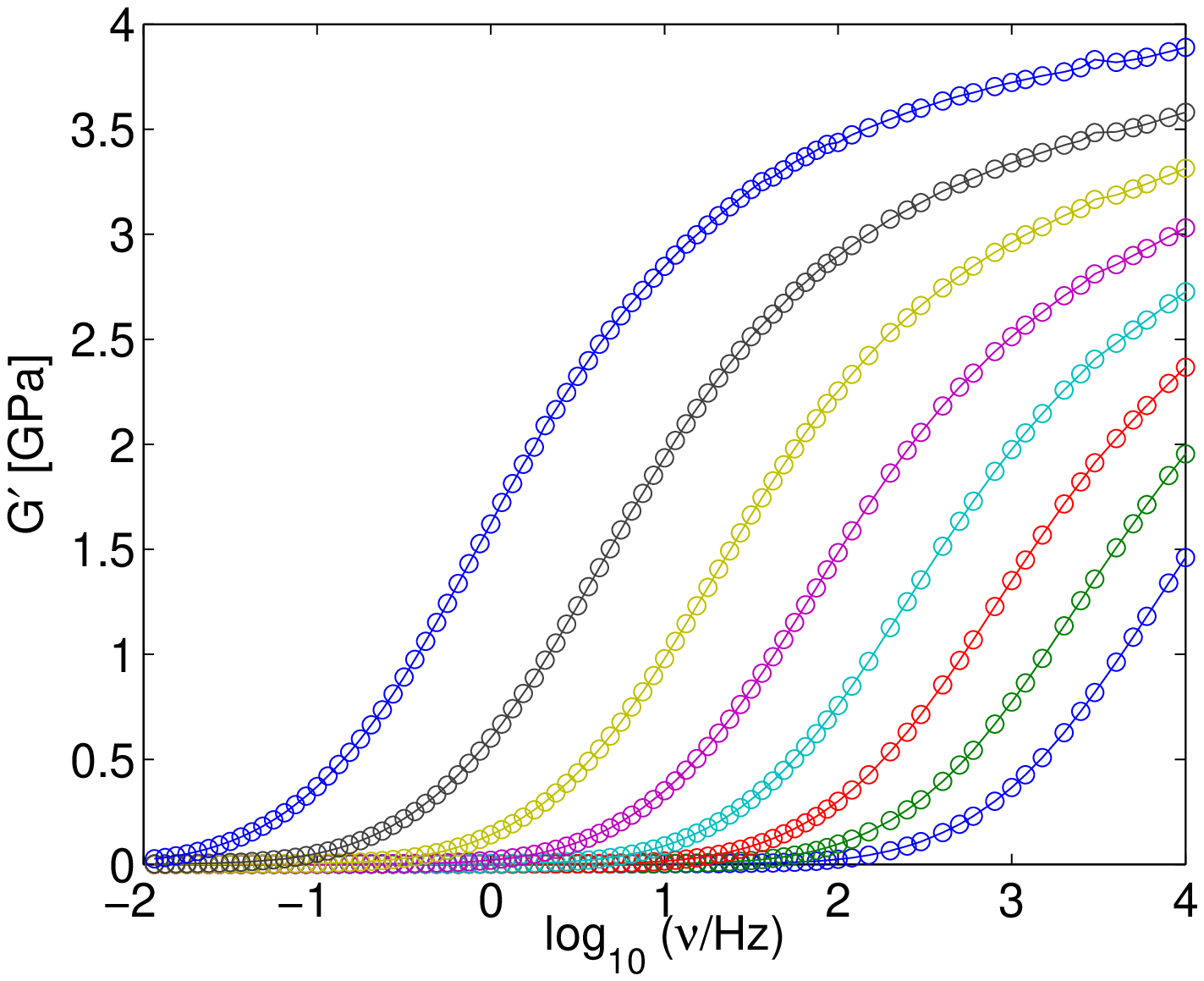,width=0.35\linewidth,clip=} &
\epsfig{file=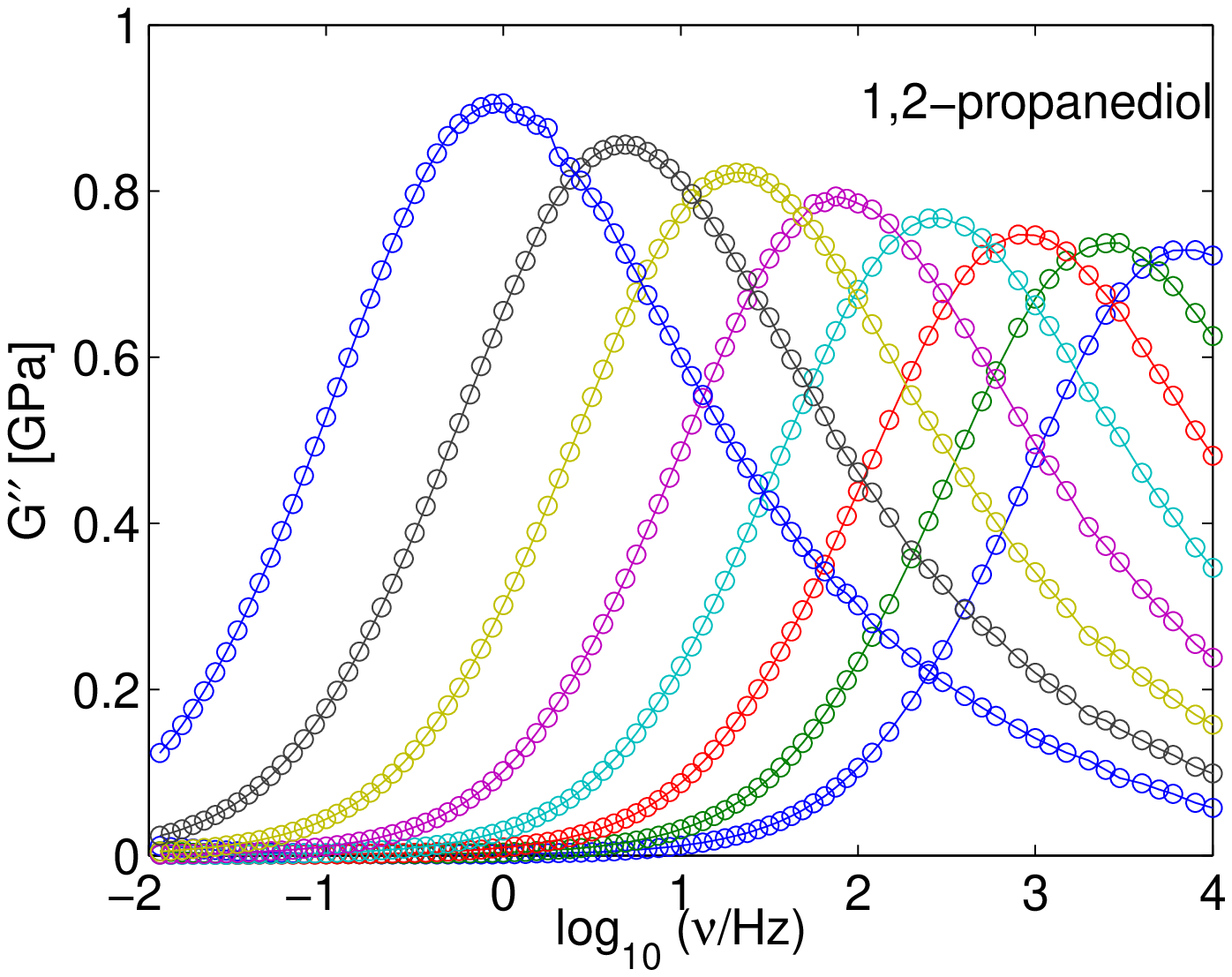,width=0.35\linewidth,clip=} \\
\epsfig{file=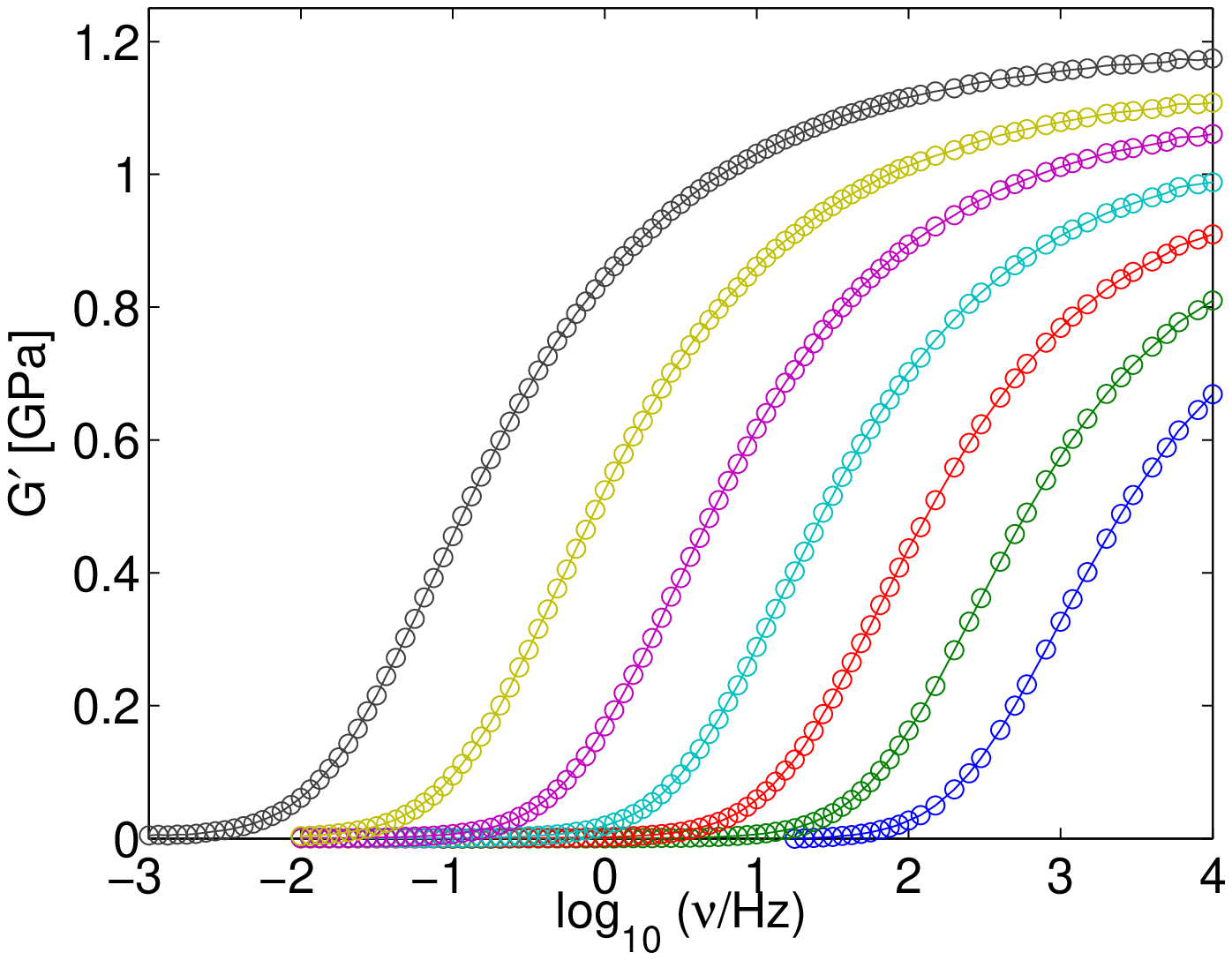,width=0.35\linewidth,clip=} &
\epsfig{file=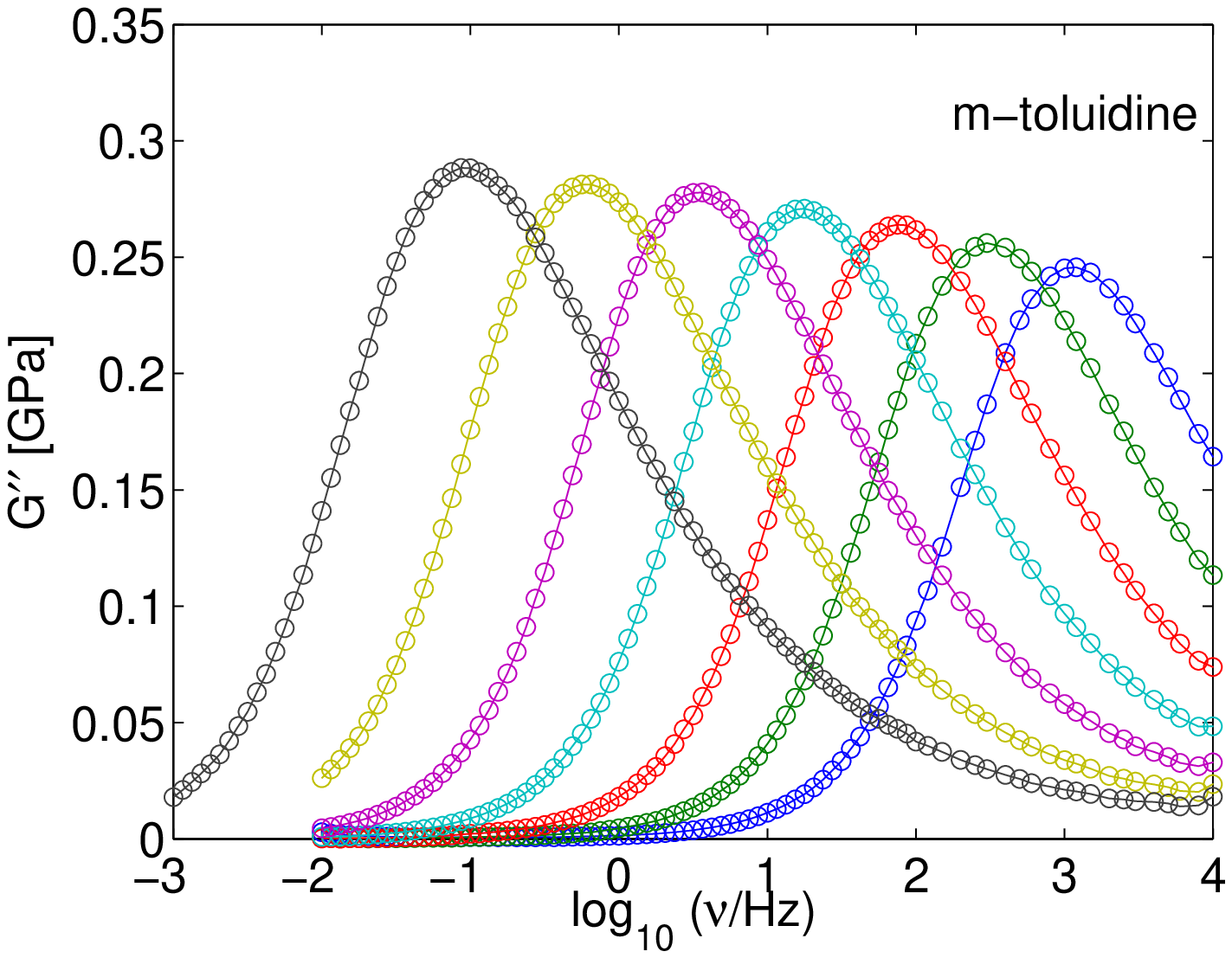,width=0.35\linewidth,clip=}
\end{tabular}
\caption{Real (left) and imaginary (right) part of the shear response of the liquids studied.  (Form top to bottom) spectra of DC705 taken at the temperatures (from right to left): 249, 246, 243, 240, 237, 234, 231 K; of
DBP at 188, 186, 184, 182, 180, 178, 176 K; of DEP at 199, 197, 195, 193, 191, 189, 187, 185, 183, 181 K; of
1,2-Propanediol at 195, 192, 189, 186, 183, 180, 177, 173, 171 K and of m-toluidine at 198,  196,  194,  192,  190, 188, 186 K.
}
\label{fig:spectra}
\end{figure*}

All mechanical spectra are reported in Figure \ref{fig:spectra}.
Here the real part $G^{\prime}$ and the imaginary part $G^{\prime\prime}$ of the complex shear
modulus $G(\nu,T)$ are presented\cite{Repos}.
The reactive and the absorbitive part of the shear-response 
of the liquids studied are measured at several temperatures.

For the shear loss peak of the DC705 there is a clearly defined 
shear $\alpha$ process that shifts to lower frequencies 
as the temperature decreases.
We have found no sign of a shear $\beta$ relaxation
in this liquid. This is also confirmed by the fact that the curve maintains the
same shape lowering the temperature as discussed below.
We will see also that the high-frequency side of the shear-mechanical 
spectrum shows a pretty 
constant slope.

The shear response of DBP clearly shows a $\beta$ relaxation.
The shape of the imaginary part of the response function is strongly 
deformed at high temperatures where the $\alpha$ process is merged with the $\beta$.
When the temperature is low enough, the high-frequency tail of the response functions
shows a pronounced increase signaling an emerging $\beta$ relaxation.
Unfortunately the $\beta$ loss peak lies at 
much higher frequencies than the upper limit of our device.
The dielectric $\beta$ process of this liquid has been observed~\cite{DBP1} to be around 
$10^6$ Hz in a temperature window similar to the one of our mechanical experiment. 
This is consistent with the fact that the $\beta$ relaxation is
outside the PSG dynamical range
since the shear-loss peak frequency is generally higher than the dielectric one~\cite{Kr&Bo},
(usually the two peaks are within the same decade).
For future studies on this liquid the upper frequency limit 
of the PSG could possibly be increased by slightly modifying the transducer geometry.

Evidence of a secondary loss peak is also found in the mechanical response of the DEP.
The dielectric $\beta$ relaxation process in DEP has been 
the subject of an accurate investigation in connection with the 
behavior of the entropy in the supercooled and in the glass state~\cite{DEP1},
raising fundamental questions about the nature of fast molecular motions
in ultra-viscous liquids. Very recently this dielectric $\beta$ process has been shown 
to be intimately related to the results of positron-annihilation lifetime spectroscopy~\cite{DEP2}.
We note that in those dielectric measurements the $\beta$ loss peak is found at about
$2\times10^4$ Hz at $T=179$ K consistent with the fact that mechanically this process 
is above the upper limit of the PSG. 
We shall see in Sec. III that when the shear loss is plotted 
on a logarithmic scale, the low-frequency tail of the $\beta$ relaxation is more clearly visible.

The spectra taken for 1,2-propanediol are characterized by an high shear-modulus and no signature of
$\beta$ relaxation is found. 
This hydrogen-bonded liquid has recently been studied in dielectric experiments and compared to polymers differing in chain length and in the number of OH group~\cite{12_1}. 
These dielectric experiments did not reveal a visible $\beta$ process, and a temperature-independent Cole-Davidson stretching parameter was found.
These findings are fully in agreement with our shear spectra since the shape of the response function is very weakly temperature dependent as it will be stressed in the following.

Rather interesting mechanical spectra are found for m-toluidine. A weak, but non-zero high-frequency tail of the loss response can be noted in Fig. \ref{fig:spectra}. This may hint to the existence of a shear $\beta$ relaxation at frequencies around 1 MHz, consistent with recent broad-band dielectric experiments performed on this liquid~\cite{m-tolu1} where a $\beta$ relaxation process could be resolved, but only in the glassy regime where the relaxation frequency reaches a value around 1 kHz at about 130 K. This could be a range in which this $\beta$ process could start to be detected mechanically\cite{prob}.

Finally we mention that we attempted to measure the frequency-resolved shear-modulus also for 1,3-propanediol and for propylene carbonate. An anomalous shear response was found for these two liquids, however, signaling a probable crystallization.

\section{III. Discussion} 
\label{sec:Disc}

\begin{figure}
\begin{center}
\includegraphics[width=8cm]{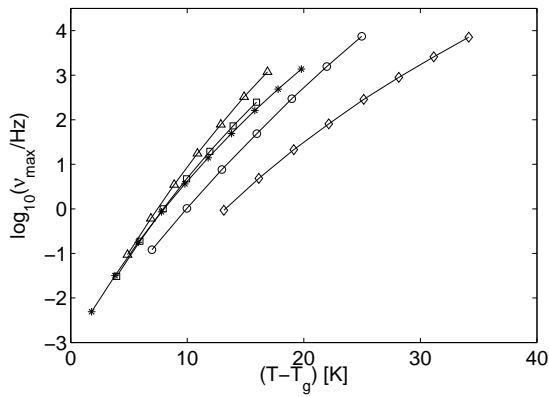}
\end{center}
\caption{Frequency of the mechanical $\alpha$ peak as a function of the temperature for DC705 ({\Large{$\circ$}}, $T_g=224.0$ K), DBP ( $\Box$, $T_g=172.3$ K), DEP ($\ast$, $T_g=179.5$ K), 1,2-Propanediol ($\diamondsuit$, $T_g=157.8$ K), m-toluidine ($\bigtriangleup$, $T_g=181$ K).}
\label{fig:fvsT}
\end{figure}

The first information that can be extracted from our measurement is the $\alpha$ process
loss peak frequency $\nu_{max}$. This has been deduced from the imaginary part through 
a simple unbiased method. We fitted the closest six points to the maximum of
$\log_{10}(G^{\prime\prime/}\mathrm{GPa})$ as a function of $\log_{10}(\nu/\mathrm{Hz})$ with a second order polynomial to identify the maximum $G^{\prime\prime}_{max}$ and the corresponding frequency $\nu_{max}$.
These loss-peak frequencies are reported in Fig. \ref{fig:fvsT} as function of the scaled temperature $(T-T_g)$.
Here we define $T_{g}$ as the temperature where the $\nu_{max}=1$ mHz. $T_g$ is identified by a linear extrapolation of the last three points of $\log_{10}(\nu/\mathrm{Hz})$ as a function of  $T$.

Time-temperature superposition (TTS) is a property of the $\alpha$ process 
appearing in the susceptibilities of some super-cooled liquids expressing the fact that the shape of the response function remains the same when the system is cooled. When TTS applies, the response is only shifting its characteristic time and its amplitude if we cool the liquid. Mathematically speaking the generic response $\chi$ can be rewritten as $\chi(\nu,T)=A(T) \Phi(\nu/\nu_{max}(T))$ where $\nu_{max}(T)$ is a characteristic frequency depending on $T$.
Having found the two parameters $G^{\prime\prime}_{max}$ and $\nu_{max}$ at every given temperature, it is rather easy to check wether TTS applies for a given liquid.

\begin{figure*}
\centering
\begin{tabular}{cc}
\epsfig{file=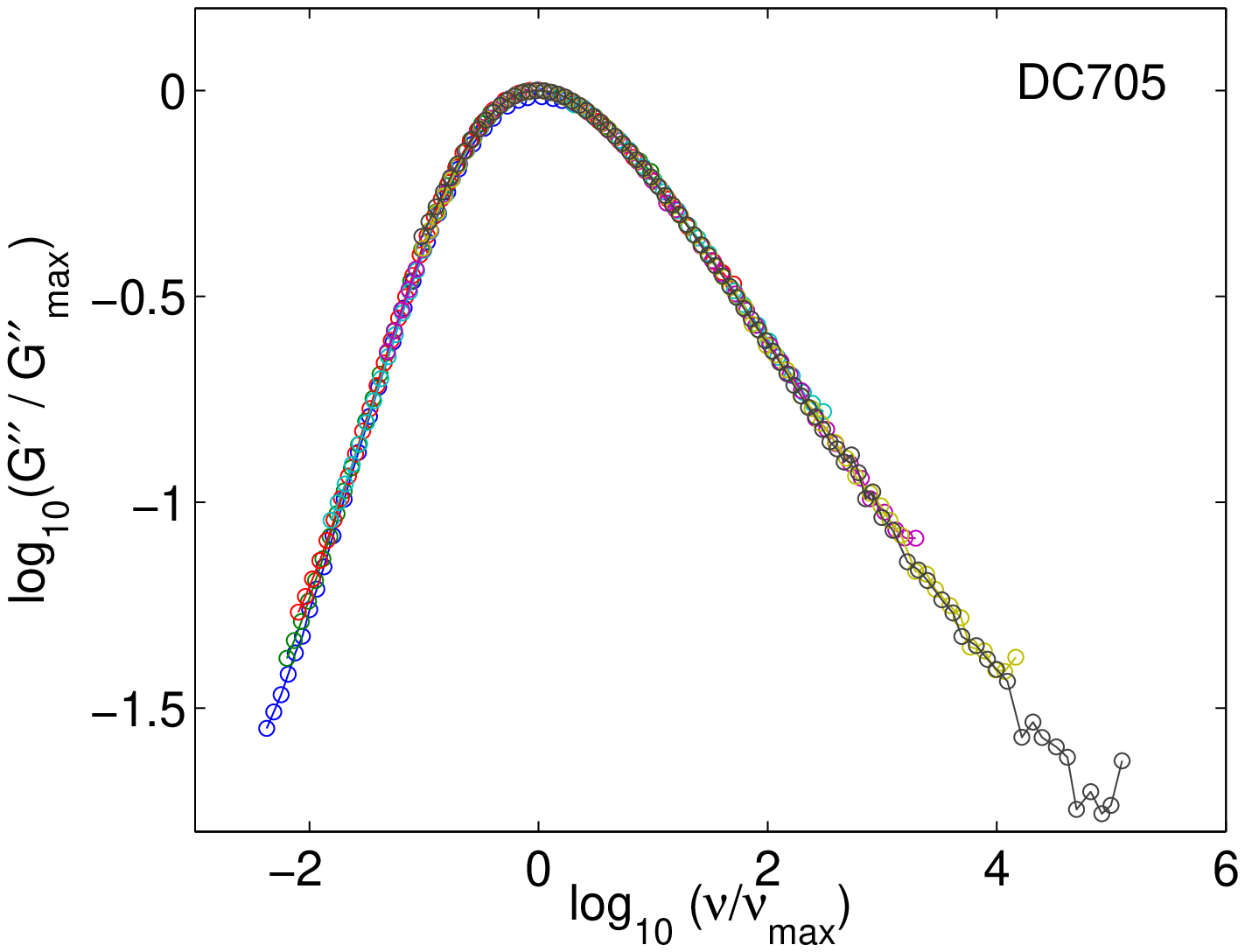,width=0.35\linewidth,clip=} &
\epsfig{file=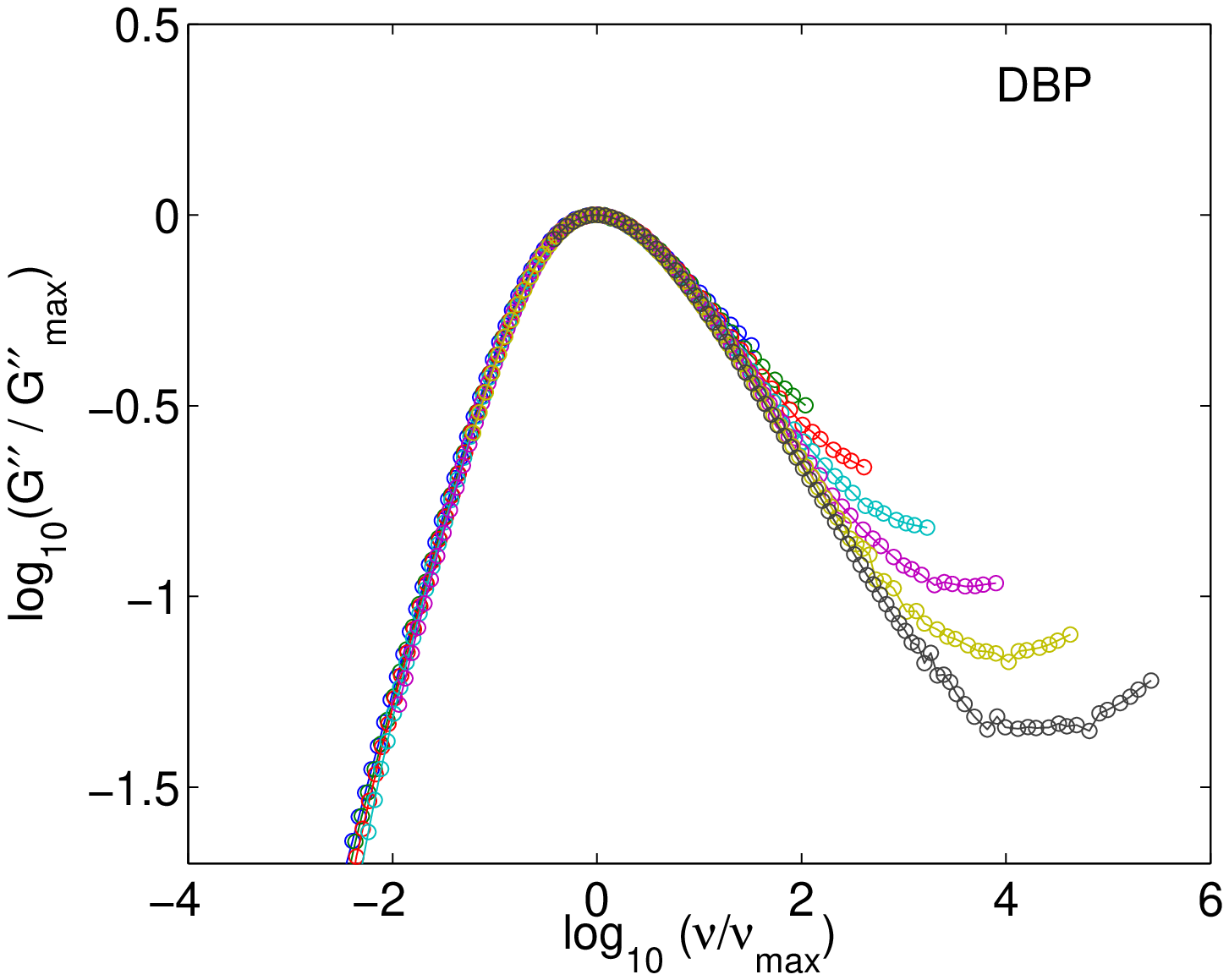,width=0.35\linewidth,clip=} \\
\epsfig{file=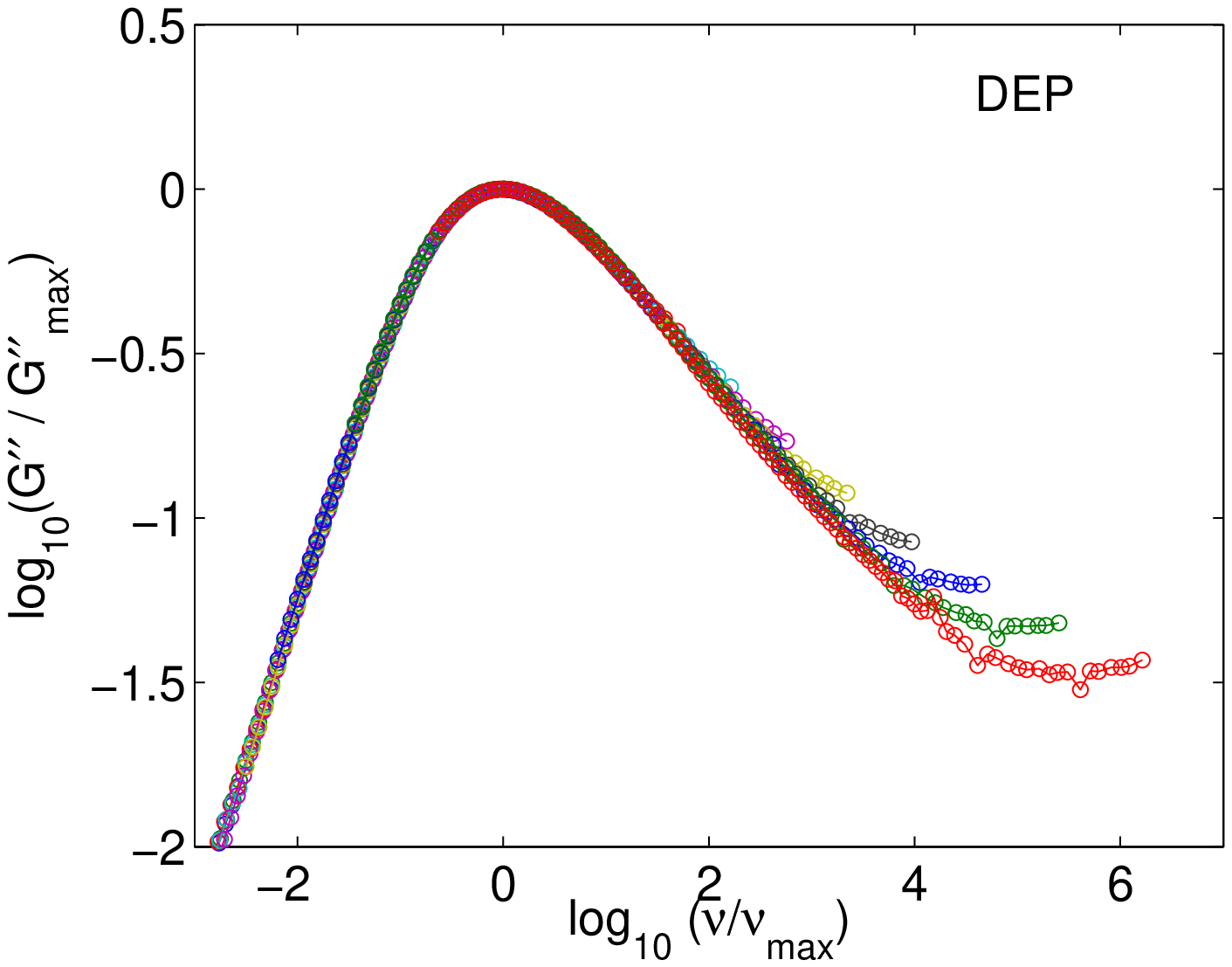,width=0.35\linewidth,clip=} &
\epsfig{file=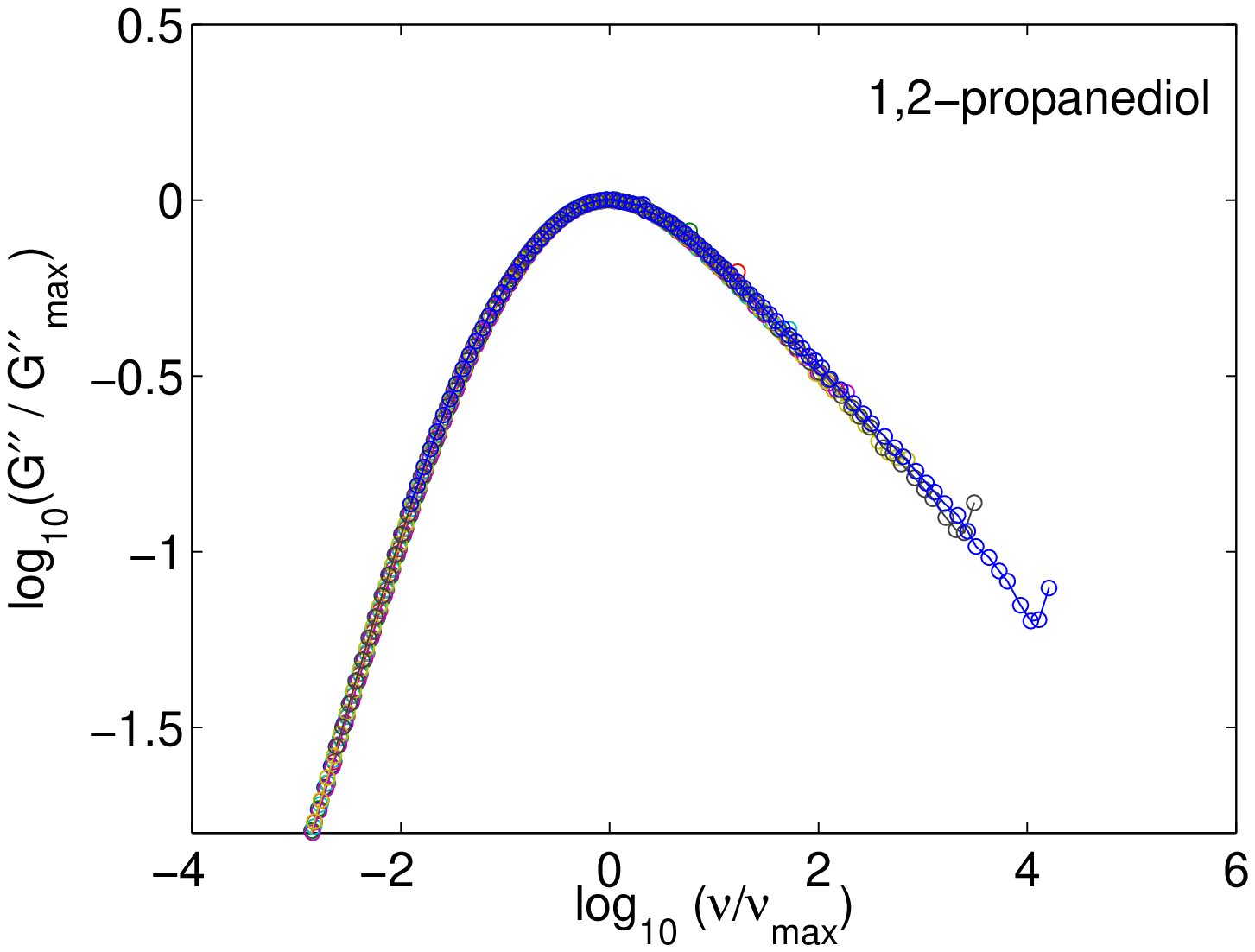,width=0.35\linewidth,clip=} \\
\epsfig{file=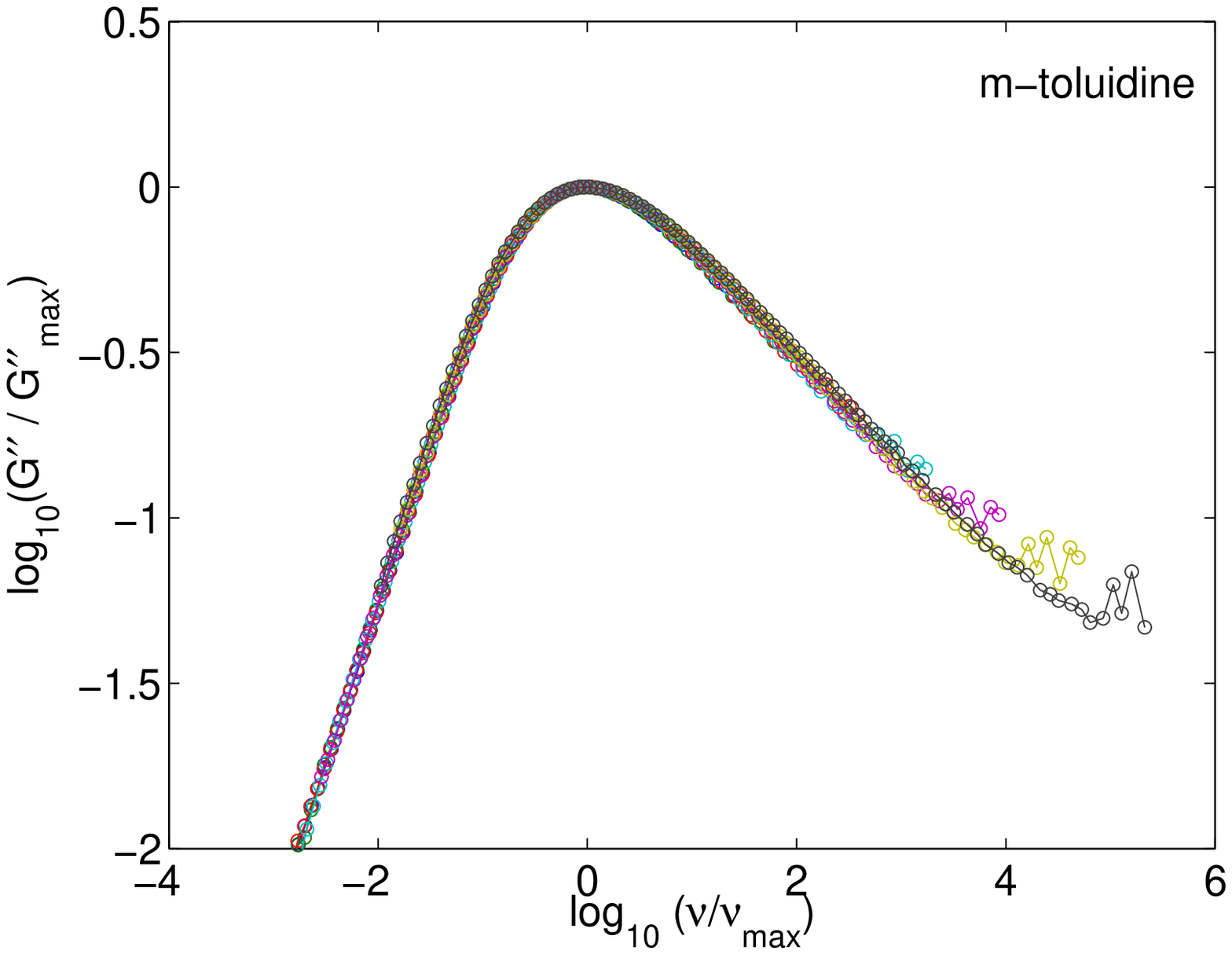,width=0.35\linewidth,clip=} &
\epsfig{file=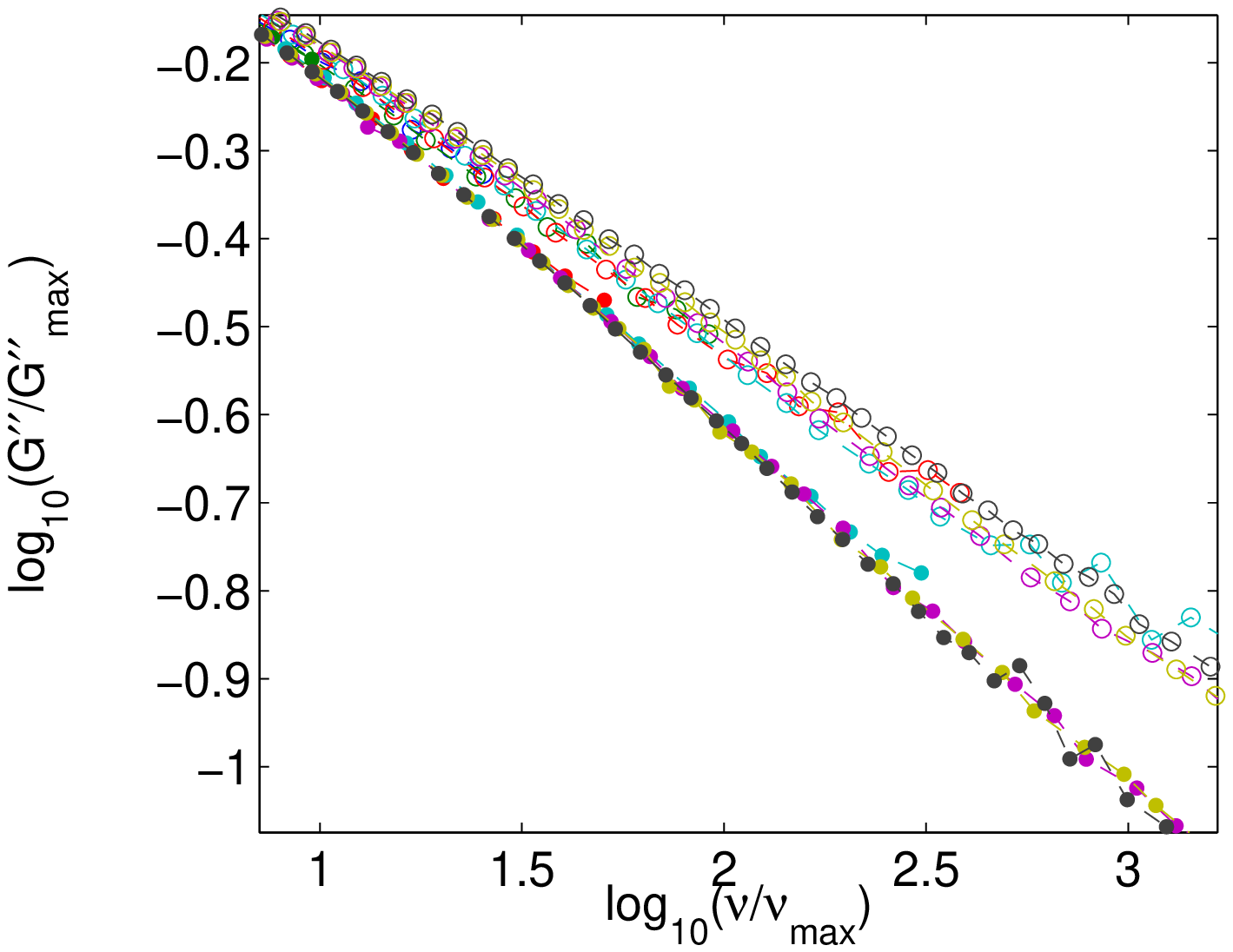,width=0.35\linewidth,clip=}
\end{tabular}
\caption{Shear loss $G^{\prime \prime}$ of liquids studied scaled by the maximum loss $G^{\prime \prime}_{max}$ as a function of the frequency $\nu$ divided by the frequency of the peak $\nu_{max}$ (see also Fig. \ref{fig:spectra}): DC705, DBP, DEP, 1,2-propanediol and m-toluidine. In the right-bottom figure we show a comparison between the scaled spectra of DC705 (full line, full symbols) and m-toluidine (dashed line, open symbols) on a smaller scale.}
\label{fig:TTS}
\end{figure*}

In Fig. \ref{fig:TTS} we report the imaginary part of the shear modulus divided by $G^{\prime\prime}_{max}$ as function of $\nu/\nu_{max}$. This plot underlines that for some of the liquids (DC705 and 1,2-propanediol) the shape of the alpha process remains constant lowering $T$. This is not the case for DBP and DEP where the curves are detaching at high frequencies. The lack of TTS is clearly related to the presence of a $\beta$ peak, indeed the logarithmic scale used in this plots makes the the low-frequency tail of the secondary process more visible. For m-toluidine the situation is ambiguous. Indeed for this liquid TTS seems to apply, but if we look at the high frequency end on a smaller scale (Fig. \ref{fig:TTS} right-bottom) there is a sensitive difference between the curves. In this figure we report also the scaled spectra of DC705 that satisfy TTS on the small scale for an easier comparison. We also note that the dielectric strength of the secondary relaxation reported for m-toluidine~\cite{m-tolu1} is rather low while in the shear-mechanical spectra it seems to be more intense in comparison with the $\alpha$ process. This is another important feature of our mechanical measurement: a weak dielectric secondary process is \emph{magnified} in the shear response~\cite{Kr&Bo,DiMarzio}.

An interesting feature of the spectra of 1,2-propanediol is underlined by the TTS plot in Fig. \ref{fig:TTS}.
The portion of the spectrum at frequencies lower than $\nu_{max}$ (i.e. at $\nu/\nu_{max} \leq 1$) is slightly deformed.
The shape of this response function is reproducible and it seems to be almost temperature-independent in the temperature range explored in our experiment.
This ``shoulder'' could be considered a peculiar feature of the mechanical relaxation of 1,2-propanediol since it is not found in dielectric spectra (see for example Ref.~\cite{12_1}).
Note that an additional process (with similar shape) close to the alpha relaxation is found in the case of tri-propylene glycol in high pressure conditions (see Ref.~\cite{12_2}).
Further studies of the dielectric relaxation of 1,2-propanediol at high pressure would be needed 
to better understand the origin of the shape of its mechanical response.

We want to stress again that, from our study, the lack of TTS seems to be intrinsically 
related to the presence of a secondary mechanical process.
All the liquids studied here that show the $\beta$ relaxation (as also confirmed by the dielectric measurements cited in Sec. II),
display a clear temperature-dependence in the shape of the mechanical response function. 
Anyway it is not generally clear how the presence of the beta relaxation is related to the nature of a specific liquid.
The understanding of the validity of TTS in some supercooled liquids requires the understanding of one of
the most puzzling phenomena of ultra-viscous liquids and glasses that is the secondary relaxation process.

\begin{figure}
\begin{center}
\includegraphics[width=8cm]{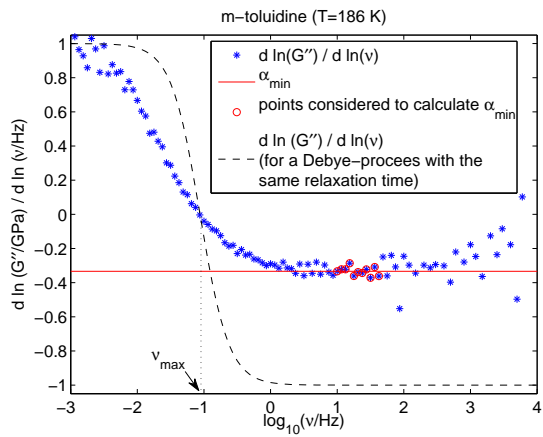}
\end{center}
\caption{Representation of the procedure used to find the minimum logarithmic slope $\alpha_{min}$.
In this figure we report the analysis of the mechanical spectrum of m-toluidine at $T=186$K.
The derivative of $\log(G''/$GPa) with respect to $\log(\nu/$Hz) is plotted as a function the frequency (stars).
Some points are selected (red circles) and their average is taken to calculate $\alpha_{min}$ (red line).
Note that a Debye-like mechanical process with the same relaxation time would have a sharper drop of the $\alpha$ parameter reaching $-1$ at high frequencies.}
\label{fig:anamin}
\end{figure}

It has been suggested from empirical observations~\cite{onehalfexp,albena} that when TTS applies, the high frequency decay of the dielectric loss is characterized by a power-law behavior $\nu^\alpha$ with exponent $\alpha=-1/2$.
A theoretical explanation for this power-law dependence of the dielectric response has been proposed assuming the dominance of long-wavelength fluctuations~\cite{onehalfthe,ome1,ome2}. We have tested this conjecture for the mechanical response finding the minimum value of the logarithmic derivative of $G^{\prime\prime}$, i.e. $\alpha_{min}=[d\ln G^{\prime\prime}/d\ln\nu]_{min}$. It has to be noted that the extraction of this information is much more complicated for the shear loss than the dielectric response. This is due to the fact that the shear response is generally affected by a higher noise level than dielectric. 
An example of the procedure followed to extract $\alpha_{min}$ is reported in Fig. \ref{fig:anamin}
for the mechanical response of m-toluidine at $T=186$K.
When $\alpha$ is plotted as a function of the frequency a minimum can be identified within some data points.
The average of these points is taken as the $\alpha_{min}$.
Note that a Debye-process would have a characteristic behavior of the logarithmic slope showing a sudden drop of $\alpha$ at the relaxation frequency $\nu_{max}$.

\begin{figure}
\begin{center}
\includegraphics[width=8cm]{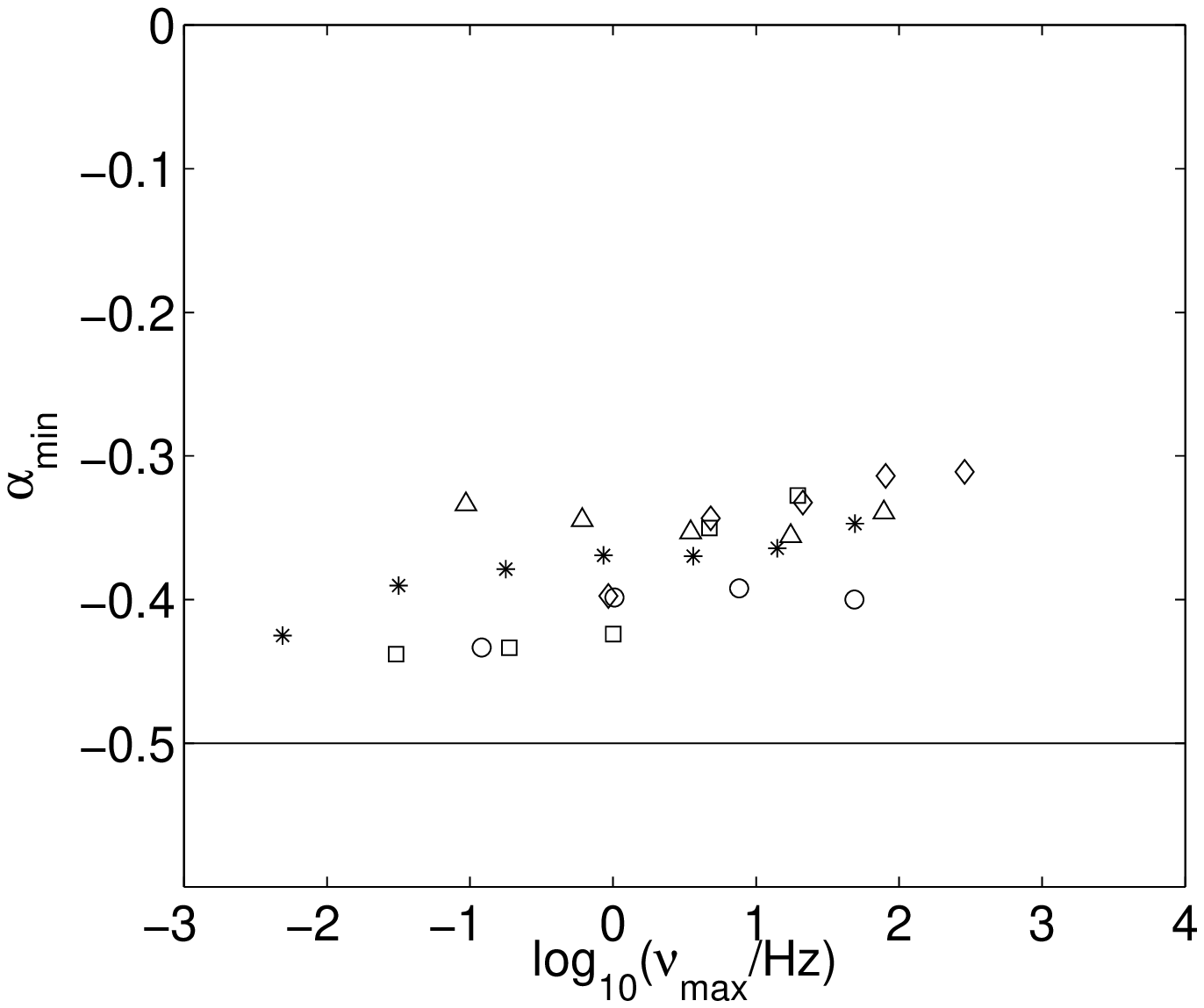}
\end{center}
\caption{Minimum value of the logarithmic slope $(\alpha=d\ln G^{\prime\prime}/d\ln\nu)$ for the liquids studied: DC705 ({\Large{$\circ$}}), DBP ( $\Box$), DEP ($\ast$), 1,2-Propanediol ($\diamondsuit$), m-toluidine ($\bigtriangleup$).}
\label{fig:amin}
\end{figure}

In Fig. \ref{fig:amin} we report the value of $\alpha_{min}$ as a function of the loss peak frequency for the temperatures where the minimum of the derivative of the logarithm was well defined. We note that even if the behavior of this quantity for the liquids studied is compatible with a limiting $\alpha=-1/2$ at low-temperatures, the data seems to lie systematically above the $-0.5$ line. We stress also that is easy to expect that an $\nu^{-1/2}$ behavior in the shear loss is disturbed more when a $\beta$ relaxation is present since its intensity is enhanced in the mechanical spectrum compared to the dielectric.

Finally we remark that the parameter $\alpha_{min}$ is not expected to monotonically decrease in a wider temperature range. Indeed at higher temperatures the shape of the shear-response should recover a simpler Debye-like shape where the stretching parameters (the KKW $\beta$ parameter for example) come close to the unity\cite{betaone}. The relaxation frequencies belonging to this high temperature regime are currently outside the dynamical window of the PSG technique\cite{dwin}.

To conclude the analysis of our data we want to present a comparison between the mechanical spectral features and a simple phenomenological model for the dynamics of supercooled liquids. The shoving model~\cite{shovin} is based on the assumption that the relaxation in the supercooled liquid takes place with a local volume increase and that the activation energy is mainly elastic energy spent to shove aside the surrounding of the rearranging molecules. As it has been reported this model can be derived estimating the barrier height in a classical energy landscape approach\cite{shovinland}. 

\begin{figure}
\begin{center}
\includegraphics[width=9.5cm]{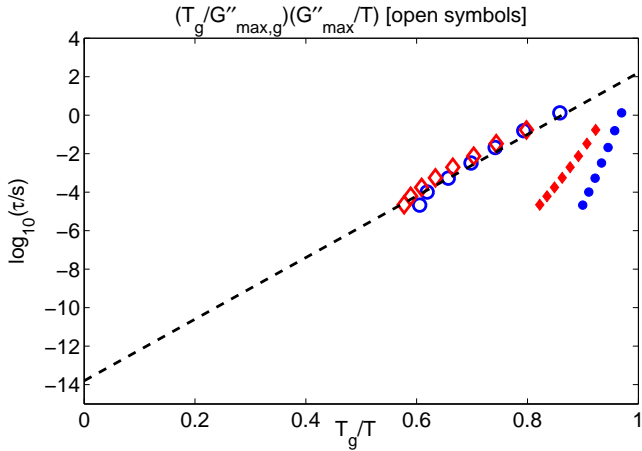}
\end{center}
\caption{Characteristic time $\tau=1/\nu_{max}$ of the shear loss as a function of the scaled temperature $T/T_g$ (full symbols) and as function of $x=(T_g/G''_{max,g})(T/G''_{max})$ (open symbols) for the liquids in which TTS applies: DC705 ({\Large{$\circ$}}, $T_g=224.0$ K) and 1,2-Propanediol ($\diamondsuit$, $T_g=157.8$ K). The dashed line is an exponential function with prefactor $\tau_0=10^{-14}$ s and taking the value $\tau= (2 \pi 10^3)^{-1} \simeq 150 $s s for x=1 (no free parameters are adjusted in this function).}
\label{fig:shovin}
\end{figure}

The shoving model predicts that the relaxation time is related to the infinite-frequency shear modulus $G_{\infty}$ by the equation: 

\begin{equation}\label{eq:S}
\ln(\tau) = \ln{\tau_0} + V_c G_{\infty}(T)/T
\end{equation}

\noindent where $V_c$ is a characteristic volume in a relaxation process ($V_c$ is assumed temperature independent) and $\tau_0^{-1}$ is the phonon-frequency  ($\tau_0 \simeq 10^{-14}$s). The determination of $G_{\infty}$ is not necessary to test the model if TTS applies, indeed if this property holds we can write 

\begin{equation}\label{eq:T}
G_{\infty}(T) \propto G^{\prime\prime}_{max}(T) 
\end{equation}

\noindent since the constant factor $A$ determining the amplitude of the complex response in $\chi(\nu,T)=A(T) \Phi(\nu/\nu_{max}(T))$ is the same for the real and the imaginary part $\Phi'$ and $\Phi''$. Note that this proportionality does not apply when TTS does not hold. 
For example in a case where the $\beta$ relaxation is present the value of $G_{\infty}$ is modified by the secondary process and a more complicated fitting procedure would be needed to estimate the value of the infinite-frequency shear-modulus. On the other hand, when TTS is satisfied, we can write the simple equation (using Eq.(\ref{eq:T}) in Eq.(\ref{eq:S})) 

\begin{equation}\label{eq:SP}
\log_{10}(\tau)= \log_{10}{\tau_0} + B G^{\prime\prime}_{max}(T)/T
\end{equation}

\noindent where $B$ is a constant factor. Eq (\ref{eq:SP}) represents the prediction of the shoving model when TTS also applies: the logarithm of the relaxation time is a linear function of the quantity $G^{\prime\prime}_{max}(T)/T$. This is tested in Fig. \ref{fig:shovin} for the liquids in which time-temperature superposition is found to hold (DC705 and 1,2-propanediol). In this figure the relaxation time (defined as $\tau=(2 \pi \nu_{max})^{-1}$) is reported as a function of $1/T$ and as a function of $G^{\prime\prime}_{max}/T$. To report all the data in the same plot the abscissa has been normalized to the unity as $T_g/T$ and as $x=(T_g/G^{\prime\prime}_{max,g})(G^{\prime\prime}_{max}/T)$. Note that here we find $G^{\prime\prime}_{max,g}$ through a linear extrapolation of the last values $G^{\prime\prime}_{max}$ to $T_g$ previously identified to give $\nu_{max}(T_g) = 10 ^{-3}$Hz (corresponding to $\tau(T_g)= (2 \pi 10^3)^{-1} \simeq 150 $s). 

The dashed line in Fig. \ref{fig:shovin} represents the shoving model prediction (no adjustable paremeters are used in this function) ending, at high temperature, at the physically reasonable prefactot $10^{-14}$s.

\section{IV. Conclusion} 
\label{sec:Conc}

We have reported shear-mechanical spectra of five glass-forming liquids close to $T_g$.
Via the PSG~\cite{RevSci} technique we have investigated the behavior of the mechanical $\alpha$ process and found evidence of the presence of a mechanical $\beta$ relaxation in dibutyl phthalate, diethyl phthalate and in m-toluidine. Time-temperature superposition for the mechanical susceptibility is found to hold for the liquids without signature of $\beta$ process (pentaphenyl trimethyl trisiloxane and 1,2-propanediol). The conjecture originally developed for the dielectric response that, when TTS applies, a $\nu^{-1/2}$ decay is found for the high frequency part of the loss is checked for the shear response. We find that the data are consistent with a limiting $\nu^{-1/2}$ low-temperature behavior although the minimum slopes are systematically higher than $-0.5$. The shoving model has been tested for the two liquids without $\beta$ relaxation finding that it well describes the experimental data.

\section{Acknowledgements}

Center for viscous liquid dynamics \emph{Glass and Time} is sponsored by The
Danish National Research Foundation (DNRF).


\begin{thebibliography}{0}
\bibitem{prob} Note that a number of problems could be present in such low-temperature measurements.
There is the possibility of damaging the piezo-discs 
because of the sandwiched glass contracting, nevertheless unknown memory effects could be induced in piezo-ceramics lowering repetitively 
their temperature when the stresses are not fully relaxed.
\bibitem{dwin} The high-frequency limit of the PSG technique (10 kHz) is set by the geometrical and material properties of the piezo-electric discs. Above this frequency the response of the measuring device is 
dominated by the intrinsic resonances of the piezo-ceramics (see Ref.[1]).
\end{thebibliography}
\end{document}